# Electric dipole in a magnetic field: some aspects of the problem

*A. P. Dmitriev, Ioffe Institute, Politekhnicheskaya 26, 194021, St. Petersburg, Russia*

In the paper some regimes of motion of an electric dipole placed in a uniform magnetic field are considered. The motion of both three-dimensional and two-dimensional dipole in the plane perpendicular to the magnetic field is studied. In the case of a two-dimensional dipole is discussed, in particular, the regime of chaotic dipole motion, which arises when, along with the magnetic field, there is an electric field rotating in the plane of motion. The motion of a three-dimensional dipole is discussed in cases that allow for analytical consideration. Finally, the same examples are used to discuss the quantum-mechanical approach to the problem.

## I. INTRODUCTION

The motion of a charged particle in an magnetic field has been studied in a large number of scientific articles from both classical and quantum points of view, while the motion of a neutral particle with a nonzero electric dipole moment is devoted to only a few papers [1-4]. Meanwhile, the theoretical study of the behavior of such a particle in an electromagnetic field is of considerable scientific and practical interest, since it can serve as a good model of polar molecules in gases and liquids, excitons in solids, and also be an element of nanomechanical devices.

In the Refs. [1-4] have shown that a classical electric dipole placed in a magnetic field oscillates (rotates) around its center of mass, which, in turn, moves along a trajectory whose shape can be quite bizarre. In Ref. [1], specific examples of the so-called trapped stats of a three-dimensional dipole in a uniform magnetic field are discussed and general considerations are given regarding the conditions for the occurrence of such states in both classical and quantum mechanics. In [2-3] the motion of a two-dimensional dipole in a plane perpendicular to the external magnetic field has been studied and some special regimes of motion, in particular trapped states, have been considered. In Ref. [4], the problem of a three-dimensional dipole in a uniform magnetic field is considered in the approximation of small oscillations of the dipole, as well as some types of its trajectories and trapped stats are discussed. There are also a number of works in which the classical dynamics of two particles interacting according to the Coulomb's law and moving in a uniform magnetic field is discussed, but they are not directly related to our topic.

In this paper we draw attention to some interesting, in our opinion, aspects of the problem that have not previously been discussed. The paper is organized as follows.

In Section II, we describe the system under study and obtain an expression for the Lagrange function in a form convenient both for further calculations and for presentation of the results. To derive such an expression, we took advantage of the fact that the Lagrange function is defined to the accuracy of the time derivative of an arbitrary function of time and coordinates.

In Section III we study some previously undisguissed regimes of motion of a two-dimensional dipole in the plane perpendicular to the magnetic field direction. In particular, we discuss the regime of chaotic dipole motion, which can arises when, along with the magnetic field, there is an electric field rotating in the plane of motion.

Section IV describes special regimes of motion of a three-dimensional dipole admitting nontrivial analytical solutions within classical mechanics, and in section V the same regimes are considered from the viewpoint of quantum mechanics. In section V the quantum version of the two-dimensional dipole problem is also discussed.

Finally, the section VI briefly describes the main results of the work.

## II. STATEMENT OF THE PROBLEM

As a simple model of a dipole, consider a system of two particles with masses $m_1$ and $m_2$ and charges $q$ and $-q$, respectively, located at a fixed distance $l$ from each other. Directing the axis $Z$ along the magnetic field and choosing a vector potential in the form $A_x = A_z = 0, A_y = Bx$, we obtain for the Lagrange function of the system

$$L = \frac{m_1 \dot{\mathbf{r}}_1^2}{2} + \frac{m_2 \dot{\mathbf{r}}_2^2}{2} + \frac{qB}{c}(\dot{y}_1 x_1 - \dot{y}_2 x_2) \tag{1}$$

We introduce the vector of the center of mass $\mathbf{r} = (x, y, z)$ and the vector of the relative position of the particles $\mathbf{l} = \mathbf{r}_1 - \mathbf{r}_2 = (l_x, l_y, l_z)$ and rewrite the expression (1) in these variables:

$$L = \frac{m(\dot{x}^2 + \dot{y}^2 + \dot{z}^2)}{2} + \frac{\mu(\dot{l}_x^2 + \dot{l}_y^2 + \dot{l}_z^2)}{2} + \frac{qB}{c}\left(x\dot{l}_y + \dot{y}l_x - \frac{\Delta m}{m}\dot{l}_y l_x\right), \tag{2}$$

$$m = m_1 + m_2, \quad \Delta m = m_1 - m_2, \quad \mu = \frac{m_1 m_2}{m_1 + m_2}$$

Let us now take advantage of the fact that the Lagrange function is defined to the accuracy of the time derivative of an arbitrary function of coordinates and time and subtract from the right-hand side of expression (2) the derivative of $(qB/c)(l_y x - \Delta m l_x l_y / 2m)$, after which we obtain:

$$L = \frac{m(\dot{x}^2 + \dot{y}^2 + \dot{z}^2)}{2} + \frac{\mu(\dot{l}_x^2 + \dot{l}_y^2 + \dot{l}_z^2)}{2} + \frac{qB}{c}\left[\dot{y}l_x - \dot{x}l_y - \frac{\Delta m}{2m}(\dot{l}_y l_x - l_y \dot{l}_x)\right]. \tag{3}$$

This function is independent of the coordinates of the center of mass, so that the corresponding momenta are integrals of motion:

$$p_x = m\dot{x} - \frac{qB}{c}l_y = \text{const}, \quad p_y = m\dot{y} + \frac{qB}{c}l_x = \text{const}, \quad p_z = m\dot{z} = \text{const} \tag{4}$$

Everywhere below, without loss of generality, we will put $p_z = 0$ and $z = 0$, i.e. we will assume that the motion of the dipole center of mass occurs in the plane $(X,Y)$.

## III. THE CASE OF A TWO-DIMENSIONAL DIPOLE

In this section, we will discuss several regimes of dipole motion, when not only its center of mass is in the plane $(X,Y)$, but also the dipole moment vector belongs to this plane, i.e. $l_z = 0$. In such a situation, instead of variables $l_x, l_y$, it is convenient to use cylindrical variables $\varphi$ and $l = \mathrm{const}$, which gives

$$L = \frac{m(\dot{x}^2 + \dot{y}^2)}{2} + \frac{I\dot{\varphi}^2}{2} + \frac{Bd}{c}(\dot{y}\cos\varphi - \dot{x}\sin\varphi), \quad I = \mu l^2, \quad d = ql, \tag{5}$$

$$p_x = m\dot{x} - \frac{Bd}{c}\sin\varphi = \mathrm{const}, \quad p_y = m\dot{y} + \frac{Bd}{c}\cos\varphi = \mathrm{const}, \quad p_\varphi = I\dot{\varphi}$$

The term in Eq. (3), which is proportional to $\dot{l}_y l_x - l_y \dot{l}_x$, is omitted in equations (5), since in this case it is equal to the full time derivative: $\dot{l}_y l_x - l_y \dot{l}_x = l^2 \dot{\varphi}$. From Eqs. (5) the expression for the Hamilton function of the dipole follows

$$H = \frac{p_x^2 + p_y^2}{2m} + \frac{p_\varphi^2}{2I} + \frac{p_x Bd}{mc}\sin\varphi - \frac{p_y Bd}{mc}\cos\varphi + \frac{B^2 d^2}{2mc^2} \tag{6}$$

and the corresponding Hamilton equations

$$\dot{p}_x = 0, \; \dot{x} = \frac{p_x}{m} + \frac{Bd}{mc}\sin\varphi, \quad \dot{p}_y = 0, \; \dot{y} = \frac{p_y}{m} - \frac{Bd}{mc}\cos\varphi, \tag{7}$$

$$\dot{p}_\varphi = -\frac{p_x Bd}{mc}\cos\varphi - \frac{p_y Bd}{mc}\sin\varphi, \quad \dot{\varphi} = \frac{p_\varphi}{I}$$

In what follows, using the independence of the impulses $p_x$ and $p_y$ from time, we direct the ordinate axis along the impulse, so that we will have $p_x = 0, \; p_y = p$, and the Hamilton function and Eqs. (4) will be written in the form

$$H = \frac{p^2}{2m} + \frac{p_\varphi^2}{2I} - \frac{pBd}{mc}\cos\varphi + \frac{B^2 d^2}{2mc^2}, \tag{8}$$

$$\dot{x} = \frac{Bd}{mc}\sin\varphi, \quad \dot{y} = \frac{p}{m} - \frac{Bd}{mc}\cos\varphi, \quad \ddot{\varphi} = -\frac{pBd}{mIc}\sin\varphi$$

where the angle $\varphi$ is measured from the new abscissa axis. The last of these equations coincides with the equation of a nonlinear mathematical pendulum with a frequency of small oscillations

$$\omega_0 = \sqrt{pBd/mIc} \tag{9}$$

The shape of the trajectory of the center of mass of the dipole depends significantly on the magnitude of the momentum $p$ and the initial value of the angular velocity $\dot{\varphi}(0) = p_\varphi(0)/I$. Below we will look at a few typical examples.

## A. The simplest possible dipole trajectories

The simplest dipole movement is zero momentum movement $p = 0$. In this case from Eqs. (8) we have (here and everywhere below we assume that at the initial moment the dipole is located at the origin of coordinates)

$$\varphi(t) = \omega t + \alpha, \quad x(t) = -\frac{\omega_c l}{\omega}\cos(\omega t + \alpha) + \frac{\omega_c l}{\omega}\cos\alpha,$$

$$y(t) = -\frac{\omega_c l}{\omega}\sin(\omega t + \alpha) + \frac{\omega_c l}{\omega}\sin\alpha, \quad \omega = \frac{p_\varphi}{I}, \quad \omega_c = \frac{qB}{mc}$$

These expressions mean that the dipole rotates around its center of mass with an arbitrary frequency $\omega$ and simultaneously rotates as a whole with the same frequency along a circle of radius $R_\omega = \omega_c l / \omega$ centered at a point $(R_\omega \cos\alpha, R_\omega \sin\alpha)$. According to Eq. (6), the dipole energy is given by the expression

$$E = \frac{p_\varphi^2}{2I} + \frac{B^2 d^2}{2mc^2},$$

where the first term on the right-hand side corresponds to the rotation of the dipole around the center of mass, and the second to the rotation of the dipole as a whole (according to formulas (8), at $p = 0$ equality $B^2 d^2 / 2mc^2 = m(\dot{x}^2 + \dot{y}^2)/2$ is true). The described movement is one of the trapped stats discussed in Refs. [2-3]. Note that at $\omega \to 0$, the radius $R_\omega$ increases indefinitely and the motion becomes uniform and rectilinear.

Another relatively simple trajectory is realized in the case of small oscillations of the dipole $\varphi \ll 1$, when, as follows from (8) $\varphi(t) = \Phi \sin(\omega_0 t + \alpha)$, which in an approximation linear in amplitude $\Phi$ gives

$$x(t) = -\frac{\omega_c \Phi}{\omega_0} l \cos(\omega_0 t + \alpha) + \frac{\omega_c \Phi}{\omega_0} l \cos\alpha, \quad y(t) = \left(\frac{p}{m} - \omega_c l\right) t \qquad (10)$$

From these expressions it can be seen that the center of mass of the dipole moves uniformly along the ordinate and performs harmonic oscillations with frequency $\omega_0$ along the abscissa. For $\omega_c \Phi / \omega_0 \gg 1$, which is equivalent,

$$Bd \gg pc \qquad (11)$$

the amplitude of these oscillations is much greater than the size of the dipole $l$.

In the case $p = m\omega_c l$, which is equivalent to equality $Bd = pc$, it follows from Eqs. (10) $y(t) = 0$, that is, the center of mass of the dipole is motionless on the ordinate and performs harmonic oscillations along the abscissa. This means we again have one of the trapped stats discussed in Refs. [2-3].

## B. Dipole motion in the vicinity of the separatrix of a nonlinear pendulum

The most interesting, in our opinion, is the situation when the energy $\varepsilon$ of oscillations (rotation) of the dipole relative to its center of mass is close to the energy corresponding to the separatrix, which delimits the regions of the phase space corresponding to oscillations ($\varepsilon < \varepsilon_s$) and to rotation ($\varepsilon > \varepsilon_s$). In both cases, if the inequality $|\varepsilon - \varepsilon_s| \ll \varepsilon_s$ is fulfilled, the dipole for a long time of the order of

$$\tau(\varepsilon) = \frac{1}{\pi \omega_0} \ln\left(\frac{32\varepsilon_s}{|\varepsilon - \varepsilon_s|}\right) \gg \frac{1}{\omega_0} \tag{12}$$

is delayed in the vicinity of the "upper" point $\varphi = \pi$ (See, for example, Ref. [5] and Fig.1).

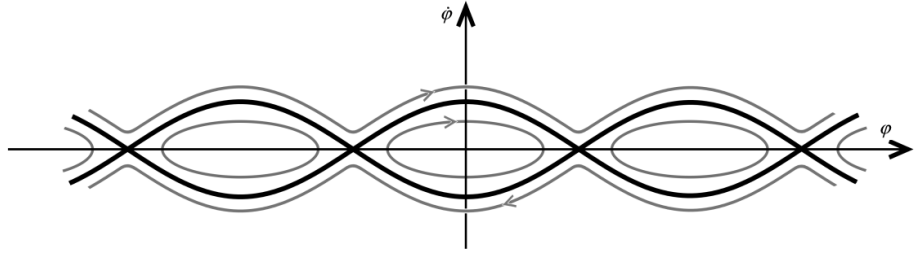

Fig. 1 The phase portrait of a nonlinear pendulum: the closed curves correspond to the oscillatory motion of the pendulum with energy $\varepsilon < \varepsilon_s$, and the open curves correspond to the rotational motion with energy $\varepsilon > \varepsilon_s$; the abscissa axis shows the pendulum rotation angle, and the ordinate axis shows its angular velocity.

The energy of oscillation/rotation of the dipole is

$$\varepsilon = \frac{I\dot\varphi^2}{2} - \frac{pBd}{mc}\cos\varphi, \tag{13}$$

whence follows

$$\varepsilon_s = \frac{pBd}{mc}. \tag{14}$$

To represent the motion of the center of mass of the dipole, we note that, according to (8),

$$x(t) = \frac{I}{p}[\dot\varphi_0 - \dot\varphi(t)], \quad \dot y(t) = \frac{p}{m} - \frac{Bd}{mc}\cos\varphi, \tag{15}$$

where $\dot{\varphi}_0 = \dot{\varphi}(0)$ is the angular velocity at the "bottom" point. From (10) we see that $\dot{\varphi}_0$, corresponding to the separatrix, is given by the expression

$$\dot{\varphi}_0 = \sqrt{\frac{2pBd}{mcI}} . \tag{16}$$

Let it be first $\varepsilon > \varepsilon_s$. In this case, the dipole quickly, in a time of the order $1/\omega_0$, rotates around the center of mass by an angle close to $\pi$, then for a long time (12) passes the "upper" point, after which, during the time of order $1/\omega_0$, it returns to the beginning. It can be seen from Eqs. (15) and Eq. (16) that simultaneously the coordinate of the dipole on the abscissa axis, within a time of the order of $1/\omega_0$, reaches its maximum value

$$x_{max} \approx \sqrt{\frac{2IBd}{mcp}} \tag{17}$$

remains close to it for a long time, after which it quickly returns to its original zero value. The quantity $\sqrt{I/m}$ is of the order of the dipole size, so it follows from Eq. (14) $x_{max} \gg l$ if inequality (11) is satisfied. In this case, the coordinate $y$ first decreases for a short time, then begins to increase, and the dipole moves along the axis ordinate for a long time uniformly with a speed approximately equal $Bd/mc$ (this section of the trajectory corresponds to uniform rectilinear motion, which was mentioned in Section IIIA). Thus, in the considered case, the trajectory of the dipole as a whole is a sequence of long sections of uniform motion along the ordinate at a distance from the abscissa, approximately equal $x_{max}$, with short-term periodic loop-like returns to $x = 0$ (Fig. 2a). A similar consideration of the case $\varepsilon > \varepsilon_s$ when the dipole performs oscillatory motions of large amplitude shows that the trajectory of the dipole is also a sequence of long links at $x = x_{max}$ with fast periodic loop-like returns to $x = 0$. The difference between this case and the previous one is that in the central part of each link, i.e. every half-period, there is an ejection along the abscissa axis to $x = 2x_{max}$, occurring during the time $\sim 1/\omega_0$ when the dipole is turned from the left to the right almost stationary position (Fig. 2b).

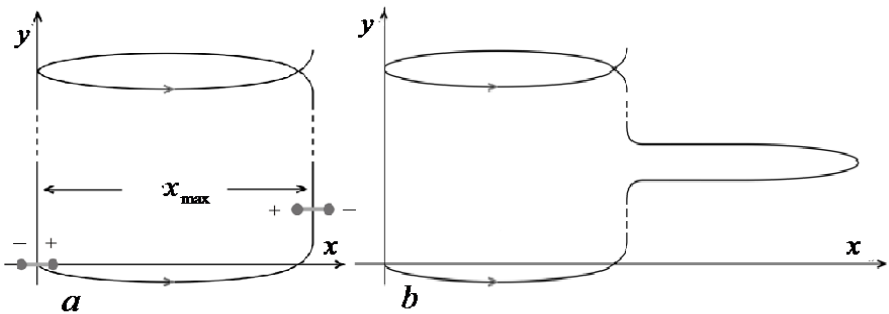

Fig. 2: *a* is trajectory of the dipole center of mass as it rotates near the separatrix; *b* is trajectory of the dipole center of mass when it vibrates near the separatrix.

We have described the motion of the dipole in the case when inequality (11) is satisfied. In the general case, as follows from Eqs. (15) and Eq. (13), the displacement of the center of mass of the dipole along the ordinate axis at $\varepsilon > \varepsilon_s$ during one revolution is given by the expression

$$\Delta y = \frac{2\sqrt{2I}\,p}{m}\left(\int_0^\pi \frac{d\varphi}{\sqrt{\varepsilon + \varepsilon_s \cos\varphi}} - \frac{Bd}{pc}\int_0^\pi \frac{\cos\varphi\, d\varphi}{\sqrt{\varepsilon + \varepsilon_s \cos\varphi}}\right), \tag{18}$$

and at $\varepsilon < \varepsilon_s$ during of one oscillation, the expression

$$\Delta y = \frac{2\sqrt{2I}\,p}{m}\left(\int_0^\Phi \frac{d\varphi}{\sqrt{\varepsilon + \varepsilon_s \cos\varphi}} - \frac{Bd}{pc}\int_0^\Phi \frac{\cos\varphi\, d\varphi}{\sqrt{\varepsilon + \varepsilon_s \cos\varphi}}\right), \tag{19}$$

where $\Phi = \Phi(\varepsilon, p)$ is the vibration amplitude. The integral $\int_0^\pi (\varepsilon + \varepsilon_s \cos\varphi)^{-1/2} \cos\varphi\, d\varphi$ is expressed in terms of complete elliptic integrals and is negative (Ref.[6]), therefore, in the case $\varepsilon > \varepsilon_s$, the coordinate increases indefinitely with time for any values of the momentum. In the case, the situation is different. Near the separatrix ($\Phi$ close to $\pi$), the integral $\int_0^\Phi (\varepsilon + \varepsilon_s \cos\varphi)^{-1/2} \cos\varphi\, d\varphi$ is also negative and the ordinate grows with time, however, with decreasing energy, it becomes positive and at some relation between $\varepsilon$ and $p$ the expression in brackets in Eq. (19) vanishes, which means the formation of a trapped state. We emphasize, however, that such states are unlikely to be of practical importance, since they arise only at a completely definite ratio between the vibration energy and the momentum of the center of mass of the dipole and are destroyed at an arbitrarily small deviation from this ratio, so that the motion becomes infinite.

C. Chaotic dipole motion

Let us now assume that an electric field rotating with frequency $\nu$ is also applied to our dipole in the plane $(X,Y)$. In this case, a term $Ed\sin(\varphi - \nu t)$ is added to the Hamilton function (6), so that it still does not depend on the coordinates of the center of mass of the dipole and the momenta conjugate to them are conserved. This again allows the ordinate axis to be directed along the momentum, which leads to the equations

$$\dot{x} = \frac{Bd}{mc}\sin\varphi, \quad \dot{y} = \frac{p}{m} - \frac{Bd}{mc}\cos\varphi, \quad \ddot{\varphi} = -\frac{pBd}{mIc}\sin\varphi - \frac{Ed}{I}\cos(\varphi - \nu t)\cdot \tag{20}$$

The last of these equations was studied in detail in the literature as applied to a mathematical pendulum and it was shown, in particular, that at a sufficiently small value of the external periodic force in the phase space around the separatrix a narrow layer is formed in which the movement of the pendulum is chaotic, i.e. practically unpredictable: trajectories with close initial conditions diverge indefinitely (see, for example, Ref. [5] and Fig. 3a). In this case, on each of these trajectories, the oscillation regime "randomly" is replaced by the rotation regime and vice versa. As applied to our problem, this leads to the fact that when the energy is close to

$\varepsilon_s$, the trajectory of the center of mass of the dipole turns out to be "random", consisting of "randomly" alternating and having "random" lengths of the links of the two types described above (Fig. 3b).

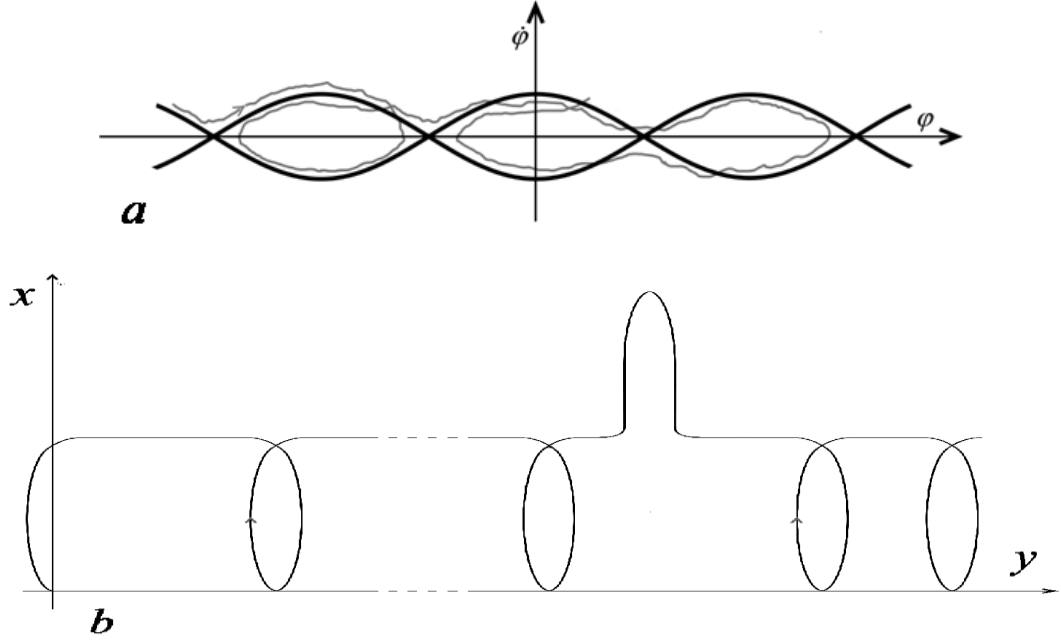

Fig. 3. *a*: phase trajectory of chaotic motion of a pendulum in the vicinity of the separatrix and *b*: the typical trajectory of the dipole center of mass.

## IV. THREE-DIMENSIONAL CASE

In this case, it is convenient to switch from variables $l_x, l_y, l_z$ to spherical variables $l = \text{const}, \theta, \varphi$ in the expression for the Lagrange function (3) :

$$L = \frac{m(\dot{x}^2 + \dot{y}^2)}{2} + \frac{I(\dot{\theta}^2 + \sin^2\theta\, \dot{\varphi}^2)}{2} + \frac{Bd}{c}(\dot{y}\cos\varphi - \dot{x}\sin\varphi)\sin\theta - \frac{Bld}{2c}\frac{\Delta m}{m}\sin^2\theta\, \dot{\varphi}, \qquad (21)$$

$$p_x = m\dot{x} - \frac{Bd}{c}\sin\varphi\sin\theta = \text{const}, \quad p_y = m\dot{y} + \frac{Bd}{c}\cos\varphi\sin\theta = \text{const},$$

$$p_\varphi = \left(I\dot{\varphi} - \frac{Bld}{2c}\frac{\Delta m}{m}\right)\sin^2\theta, \quad p_\theta = I\dot{\theta}$$

From Eqs. (21) we obtain the Hamilton function

$$H = \frac{\mathbf{p}^2}{2m} + \frac{p_\theta^2}{2I} + \frac{p_\varphi^2}{2I\sin^2\theta} + \frac{Bld}{2Ic}\frac{\Delta m}{m}p_\varphi + \frac{Bd}{mc}(p_x\sin\varphi - p_y\cos\varphi)\sin\theta + \frac{B^2 d^2}{8\mu c^2}\sin^2\theta \qquad (22)$$

and the equations for the angular variables

$$\frac{d}{dt}\left[\left(I\dot{\varphi}-\frac{Bld}{2c}\frac{\Delta m}{m}\right)\sin^2\theta\right]+\frac{Bd}{mcI}(p_x\cos\varphi+p_y\sin\varphi)\sin\theta=0$$

$$I\ddot{\theta}+\cos\theta\left[\sin\theta\left(\frac{B^2d^2}{mc^2}-I\dot{\varphi}^2\right)+\frac{Bd}{mc}(p_x\sin\varphi-p_y\cos\varphi)+\frac{Bld}{c}\frac{\Delta m}{m}\dot{\varphi}\sin\theta\right]=0$$

(23)

In the general case, these equations cannot be solved, but when $p_x = p_y = 0$ they can be analyzed analytically. Indeed, in this case it follows from the first equation

$$\sin^2\theta\left(I\dot{\varphi}-\frac{Bld}{2c}\frac{\Delta m}{m}\right)=p_\varphi=\text{const},\tag{24}$$

after which the second equation takes the form

$$I\ddot{\theta}+\sin 2\theta\frac{B^2d^2}{8\mu c^2}-\frac{p_\varphi^2\cos\theta}{I\sin^3\theta}=0.\tag{25}$$

The energy of dipole vibrations, as seen in Eq. (22), is

$$\varepsilon=\frac{p_\theta^2}{2I}+\frac{p_\varphi^2}{2I\sin^2\theta}+\frac{B^2d^2}{8\mu c^2}\sin^2\theta+\frac{Bd\Delta ml}{2mIc}p_\varphi\tag{26a}$$

and the function

$$U(\theta)=\frac{p_\varphi^2}{2I\sin^2\theta}+\frac{B^2d^2}{8\mu c^2}\sin^2\theta+\frac{Bd\Delta ml}{2mIc}p_\varphi\tag{26b}$$

plays the role of potential energy for the degree of freedom $\theta$. Using Eq. (26a) and formulas (21), the problem can be studied analytically, but we will limit ourselves here to a qualitative consideration. Function is symmetric with respect to the point $\theta=\pi/2$, has two minima at $B^2d^2l^2/4c^2>p_\varphi^2\neq 0$ and one minimum when the inverse inequality is fulfilled (the case $p_\varphi=0$ requires a separate consideration and we will not dwell on it). In the first case, at $\varepsilon<U(0)$, the dipole oscillates within one of the wells, being, on average, oriented at a certain acute (obtuse) angle to the direction of the magnetic field. On the contrary, at $\varepsilon>U(0)$, as well as in the case $B^2d^2l^2/4c^2<p_\varphi^2$, the dipole oscillates about the point of symmetry of the potential and is, on average, directed perpendicular to the magnetic field. It can be seen from (24) that in all cases the azimuthal angle $\varphi$ is a monotonic function of time, so that simultaneously with oscillations about the axis $Z$ (nutation) and the movement of the center of mass along the Cartesian axes, the dipole rotates around the axis $Z$ (precession), and the periods of nutation and precession do not coincide. In contrast to the case of a two-dimensional dipole considered above, when there were only trapped stats for $p_x=p_y=0$, under the same condition in the three-dimensional case such states apparently do not arise, since for this at the same energy in the limit $t\to\infty$ two quantities

$$\Delta x = \frac{Bd}{mc}\int_0^t \sin\varphi \sin\theta dt', \quad \Delta y = -\frac{Bd}{c}\int_0^t \cos\varphi \sin\theta dt'$$

must simultaneously remain finite.

Now let us briefly discuss the regime of motion of the dipole considered in [1]. Namely, in this paper the situation was studied when the dipole vibrates (rotates) in the plane $(Y,Z)$, so that $l_x = 0$, and at the same time can move along the Cartesian axes $X$ and $Y$. In this case, it is convenient to switch to cylindrical variables by choosing as the polar axis the axis of $X$. This will give

$$L = \frac{m(\dot{x}^2 + \dot{y}^2)}{2} + \frac{I\dot{\varphi}^2}{2} - \frac{Bd}{c}\dot{x}\cos\varphi, \quad p_\varphi = I\dot{\varphi},$$

$$p_x = m\dot{x} - \frac{Bd}{c}\cos\varphi = \text{const}, \quad p_y = m\dot{y} = \text{const},$$

$$H = \frac{(p_x^2 + p_y^2)}{2m} + \frac{p_\varphi^2}{2I} + U(\varphi), \quad U(\varphi) = \frac{B^2 d^2}{2mc^2}\cos^2\varphi + \frac{p_x Bd}{mc}\cos\varphi$$

(27)

In Ref. [1] the case $p_y = 0$ when the dipole center of mass moves only along the abscissa axis was considered in detail and the conditions of trapped stats were studied. Here we note only that the analysis performed in Section III for the case of a two-dimensional dipole - in particular, the analysis of its motion in the vicinity of the separatrix, both regular and chaotic - can also be done in this case without significant differences.

## V. QUANTUM MECHANICAL CONSIDERATION

In this section, we briefly discuss the quantum version of the problem. According to (6), in the two-dimensional case, the Schrödinger equation has the form

$$\left(\frac{p_x^2 + p_y^2}{2m} + \frac{p_\varphi^2}{2I} + \frac{p_x Bd}{mc}\sin\varphi - \frac{p_y Bd}{mc}\cos\varphi + \frac{B^2 d^2}{2mc^2}\right)\psi = E\psi,$$

(28)

$$p_x = -i\hbar\frac{\partial}{\partial x}, \quad p_y = -i\hbar\frac{\partial}{\partial y}, \quad p_\varphi = -i\hbar\frac{\partial}{\partial \varphi}$$

Substituting here $\psi(\mathbf{r},\varphi) = \exp(i\,\mathbf{pr}/\hbar)\chi(\varphi)$, we get

$$\left(-\frac{\hbar^2}{2I}\frac{\partial^2}{\partial \varphi^2} + \frac{p_x Bd}{mc}\sin\varphi - \frac{p_y Bd}{mc}\cos\varphi\right)\chi = \varepsilon_n \chi,$$

(29)

$$\chi(\varphi + 2\pi) = \chi(\varphi), \quad E = \frac{p^2}{2m} + \frac{B^2 d^2}{2mc^2} + \varepsilon_n$$

The first two terms in expression (29) for the total energy $E$ correspond to the motion of the center of mass of the dipole, and the quantities $\varepsilon_n$ are the energies of its oscillations (rotation) relative to the center of mass. By directing the ordinate along the momentum, we have

$$\left(-\frac{\hbar^2}{2I}\frac{\partial^2}{\partial\varphi^2} - \frac{pBd}{mc}\cos\varphi\right)\chi = \varepsilon_n\chi. \tag{30}$$

The solutions to this equation have been studied in detail in the literature and we will not dwell on their analysis. Note, however, that at $p=0$, equation (30) is reduced to the Schrödinger equation for a rotator with energy levels $\varepsilon_n = \hbar^2 n^2/2I$, while the rotation energy of the dipole as a whole is not quantized despite the fact that the corresponding classical trajectory is closed. Moreover, the circular motion of the center of mass of the dipole in classical mechanics in the quantum description of the problem does not manifest itself in any way in the structure of the eigenfunctions of the system. This is a general property: in quantum mechanics, the motion of the dipole center of mass is answered by plane waves, which do not "know" anything either about complex classical trajectories or classical traps. The absence of any "traces" of these latter in the wave functions of stationary states should not be surprising, given that in our problem, trapped stats arise only for strictly defined ratios between the vibration energy and the momentum of the center of mass. Of course, using Hamiltonian (28), one can write the Heisenberg equations, which formally coincide with the Hamilton equations (4), but no matter what wave function localized in a small region we take, the uncertainties of the coordinates of the center of mass will increase indefinitely with time. A detailed discussion of the problem of trapped stats in classical and quantum mechanics can be found, as already noted, in [1].

In the three-dimensional case, we have

$$\left[\begin{array}{c}-\dfrac{\hbar^2}{2m}\Delta - \dfrac{\hbar^2}{2I}\dfrac{\partial^2}{\partial\theta^2} - \dfrac{\hbar^2}{2I\sin^2\theta}\dfrac{\partial^2}{\partial\varphi^2} - i\hbar\dfrac{Bld}{2Ic}\dfrac{\Delta m}{m}\dfrac{\partial}{\partial\varphi} \\ +\dfrac{B^2 d^2}{8\mu c^2}\sin^2\theta - i\dfrac{\hbar Bd}{mc}\left(\sin\varphi\dfrac{\partial}{\partial x} - \cos\varphi\dfrac{\partial}{\partial y}\right)\sin\theta\end{array}\right]\psi = E\psi. \tag{31}$$

Substituting here $\psi(\mathbf{r},\varphi) = \exp(i\,\mathbf{pr}/\hbar)\Psi(\varphi,\theta)$, we get

$$\left[\begin{array}{c}\dfrac{\mathbf{p}^2}{2m} - \dfrac{\hbar^2}{2I}\dfrac{\partial^2}{\partial\theta^2} - \dfrac{\hbar^2}{2I\sin^2\theta}\dfrac{\partial^2}{\partial\varphi^2} - i\hbar\dfrac{Bld}{2Ic}\dfrac{\Delta m}{m}\dfrac{\partial}{\partial\varphi} \\ +\dfrac{B^2 d^2}{8\mu c^2}\sin^2\theta + \dfrac{Bd}{mc}\left(p_x\sin\varphi - p_y\cos\varphi\right)\sin\theta\end{array}\right]\Psi = E\Psi. \tag{32}$$

As in the classical problem, we restrict ourselves to considering the case $p_x = p_y = 0$. In this case, the Hamiltonian does not depend on the azimuthal angle $\varphi$ and we can put $\Psi(\varphi,\theta) = \exp(iM\varphi)\chi(\theta)$, after which equation (32) is transformed to the form

$$\left[-\frac{\hbar^2}{2I}\frac{\partial^2}{\partial\theta^2} + \frac{\hbar^2 M^2}{2I\sin^2\theta} - \frac{B^2 d^2}{8\mu c^2}\cos 2\theta\right]\chi = \varepsilon_n\chi, \quad \varepsilon_n = E - \frac{p^2}{2m} - \frac{\hbar M Bld}{2Ic}\frac{\Delta m}{m} - \frac{B^2 d^2}{8\mu c^2}, \tag{33}$$

i.e. to the Schrödinger equation for a particle in a symmetric potential with infinitely high "walls" at the edges of the interval $(0,\pi)$, so that $\chi(0) = \chi(\pi) = 0$. Below we will discuss the most interesting regime $B^2d^2/mc^2 \gg \hbar^2 M^2/I \neq 0$, when the potential is double-well with sufficiently deep wells, which makes it possible to use the semiclassical approximation. A similar problem has been discussed many times in the literature (see, for example, Ref. [7]) and it was shown that when the quasi-classical condition is satisfied, the doubly degenerate level $\varepsilon$ corresponding to the localization of a particle in the left or right well, due to the possibility of tunneling under the barrier separating the wells, splits into two closely related levels with energies $\varepsilon_d$ and $\varepsilon_u = \varepsilon_d + \Delta$, while the magnitude of the splitting $\Delta$ is proportional to the tunneling exponent. The wave function of the lower level is given by an even combination of functions in the wells, and the upper one is given by an odd one. From such a structure of wave functions it immediately follows that if at the initial moment the particle is in one of the wells, then after a long time $\tau_\Delta = \pi\hbar/\Delta$ it passes into another well and then oscillates between the wells with a period $\tau_\Delta$. As applied to our problem, this means that a dipole, initially oscillating about the mean position $\overline{\theta}$, jumps over a long time $\tau_\Delta$ into the state of oscillations about the position $\pi - \overline{\theta}$ and then jumps between these states with a period $\tau_\Delta$. Finally, let us note that in stationary states the dipole is "smeared" with the same probability between two vibration states, so that the quantum-mechanical mean value of the angle $\theta$ is equal to $\pi/2$.

In the second of the regimes of motion of the three-dimensional dipole considered in Section IV, the Schrödinger equation, according to equations (27), has the form

$$\left[-\frac{\hbar^2}{2I}\frac{\partial^2}{\partial \varphi^2} + \frac{B^2 d^2}{4mc^2}\cos 2\varphi + \frac{p_x B d}{mc}\cos\varphi\right]\chi = \varepsilon_n \chi, \quad \varepsilon_n = E - \frac{(p_x^2 + p_y^2)}{2m} - \frac{B^2 d^2}{4mc^2}.$$

It is easy to see that under the condition $|p_x|c < Bd$ the potential is double-well, and the above reasoning concerning the tunneling of the dipole between the states of oscillations in the wells, practically unchanged, can be carried over to this case as well.

## VI. CONCLUSIONS

The paper considers a number of previously not discussed problems concerning the motion in a magnetic field of a neutral particle with a nonzero electric dipole moment. The motion of a two-dimensional dipole in the plane perpendicular to the magnetic field in the situation when the energy of its rotation (oscillation) relative to the center of mass is close to its value on the separatrix separating the rotational and oscillatory regimes is studied. It is shown that when a rotating weak electric field is applied to such a dipole, the motion of its center of mass becomes chaotic. The three-dimensional dipole problem is discussed in particular cases allowing for an analytical solution. We consider in detail a case when motion of a dipole relative to the center of mass is described by an equation with a symmetric two-point effective potential so that the dipole can perform vibrations in one or in another well. Finally, we discuss a quantum-mechanical approach to the problem for both two- and three-dimensional dipoles using a number of examples. In particular, we consider the case of a double-well potential when a dipole can tunnel between the wells, periodically changing the regime of oscillations.